\documentclass[a4paper,11pt]{article}
\usepackage{pos}

\title{Discussion about a Standard Definition of the Signal-to-Noise Ratio (SNR) for Radio Signals of ultra-high-energy Particles}
\ShortTitle{Signal-to-Noise Ratio (SNR)}

\author*[a,b]{Frank~G.~Schr\"oder}
\author[c]{Amy~L.~Connolly}
\author[b]{Tim~Huege}
\author[a]{Abdul~Rehman}

\affiliation[a]{Bartol Research Institute, Department of Physics and Astronomy, University of Delaware, Sharp Lab, 104 The Green, Newark, DE 19716, USA}
\affiliation[b]{Institute for Astroparticle Physics (IAP), Karlsruhe Institute of Technology (KIT), Postfach 3640, 76021 Karlsruhe, Germany}
\affiliation[c]{Center for Cosmology and Astro-Particle Physics, Ohio State University, 191 West Woodruff Avenue, Columbus, OH 43210, USA}

\emailAdd{fgs@udel.edu}
\emailAdd{connolly@physics.osu.edu}
\emailAdd{tim.huege@kit.edu}
\emailAdd{arehman@udel.edu}

\abstract{Signal-to-noise ratios are a widely used concept for astroparticle radio detectors, such as air-shower radio arrays for cosmic-ray measurements or detectors searching for radio signals induced by neutrino interactions in ice.
Nonetheless, no common standards or methods are established for the determination of the signal-to-noise ratio: values cannot be compared between experiments, and for the same signal and noise, various methods differ by large factors on the signal-to-noise ratio.
This was the motivation to discuss a community-specific standardization at the ARENA conference 2022. 
No agreement on a common method to calculate signal-to-noise ratios was reached, however, awareness was raised that signal-to-noise ratios need to be well defined in any publications.
This includes providing sufficient information on the procedure used to determine the signal-to-noise ratio, in addition to simply stating the formula. Even when using the same definition of the signal-to-noise ratio, there is still a significant dependence on the procedure of calculation, e.g., the signal-to-noise ratio of waveforms containing only background can vary significantly depending on the size of the time interval used as signal search window.
To facilitate the interpretation of any signal-to-noise ratios in a specific study, the recommendation is to also state the mean value of the signal-to-noise ratio that the used method yields when applied to noise used in the study, e.g., the radio background measured by the corresponding experiment.}

\FullConference{%
  9th International Workshop on Acoustic and Radio EeV Neutrino Detection Activities - ARENA2022\\
  7-10 June 2022\\
  Santiago de Compostela, Spain}

\begin{document}
\maketitle

\section{Introduction}
Identifying and measuring radio signals of ultra-high-energy particles is a common challenge in all kinds of astroparticle radio detectors \cite{Huege:2016veh,Schroder:2016hrv,Connolly:2016pqr,BrayReview2016}, be it air-shower detection, the search for neutrino signals in the ice, or radio observatories looking for particle-induced radio pulses from the moon. 
The noise or background\footnote{In this proceeding, we use the words 'noise' and 'background' interchangeably, as it is often done in our field.} can be of thermal, Galactic, anthropogenic or any other origin, and depends on the experiment or model (in case of some simulation studies).
Although the challenge is common, the community uses a wide variety of definitions for the signal-to-noise ratio (SNR). 
These are either defined as amplitude ratio, which for coherent radio emission is roughly proportional to the energy of the primary particle, or as power ratios, which scale approximately quadratically with the primary energy. 
Another complication is that different experiments measure the 'signal' in different ways, and often noise is measured in a different way than the signal is. 
This typically leads to values larger than one when determining the SNR value of pure noise of whatever experiment or model, instead of the naive expectation of one (or zero in cases when the noise is subtracted from the signal).
However, that SNR value of pure noise differs drastically between experiments and, even for the very same data, depends on the procedure, such as the lengths of signal and noise intervals. 
These issues make signal-to-noise ratios incomparable between experiments and has led to misunderstandings between people from different collaborations. 

As a first step to overcome this unsatisfactory situation, we have started a discussion at the ARENA conference about a possible standard definition of the signal-to-noise ratio (SNR) for our field.
Although we could not reach a consensus, this discussion raised the awareness and brought up additional aspects. 
Despite of not yet having agreed on a common standard, everybody agreed that it is important in any publication to clearly state how the SNR is determined, which can be done by referencing another publication, and to provide the mean value of the SNR of pure background as an orientation.

\section{Aspects Considered for the Standardization of the Signal-to-Noise Ratio (SNR)}
This section summarizes the ideas presented and discussed at the ARENA workshop. 
The aspects considered for a possible standardization should also be seen as a guidance of what needs to be included in a publication to sufficiently define the SNR used in a particular analysis.

\subsection{Amplitude or Power}
Radio signals of particle cascades can be expressed in amplitude-like units (e.g., electric field strength, voltage in an antenna channel, or maximum of the Hilbert envelope of the voltage/field-strength trace), or in power-like units (e.g., energy fluence, integral of the power in the pulse, or peak power of the pulse). 
For consistency, it is highly recommended to make the same choice for both signal and noise, i.e, to measure both noise and signal in amplitude-like units, or to measure both in power-like units. 
Since the power and energy of the radio signal scale quadratically with the amplitude, and the amplitude scales approximately linearly with the energy of the electromagnetic shower component, the SNR in amplitude-like definitions increases approximately linearly with the energy of the particle cascade, whereas the SNR in power-like definitions increases approximately quadratically with the energy of the particle cascade (and thus with the energy of the primary particle initiating the cascade). 
As there seems to be a strong correlation between the results of several SNR methods \cite{Schroder:2010ffv,Glaser:2017ctn}, at least it is possible to convert between different SNR values, however, it is important to clearly state whether an amplitude- or power-like SNR is being used.

There are several advantages of using power-like units for the SNR, the main one being the ease of subtracting noise from measured signals, at least for relatively large signals. 
For amplitude-like definitions, it is not clear how to subtract noise as the interference with the signal can be both constructive or destructive. 
Whether or not an increase and decrease of the signal are equally likely due to the interference depends on the details of the procedure how the signal is found. 
In most cases, there will be a net bias changing as a function of the SNR and the signal strength needs to be corrected correspondingly.

Due to the principle of energy conservation, the situation is simpler for power-like definitions. It is usually straightforward to subtract the power of noise from the power of the measured signal. Although the possibilities of constructive and destructive interference will cause a variation of the signal due to noise, that effect is reduced when averaging the power over a sufficiently long time interval. At least on average, the power measured in the signal time interval will be the sum of the noise and signal powers. Nonetheless, even when using power-like definitions, it has to be carefully checked whether there are any biases remaining, in particular for signals close to the noise level.

Other minor advantages of using power-like definitions are an easy interpretation of the signal due to the concept of the energy fluence, an area-integration of which yields the radiation energy emitted by a shower \cite{PierreAuger:2016vya}, and the easy convertibility of the SNR to dB.
Stating the SNR in dB may facilitate communication with engineers, but it also comes with a risk because a certain SNR in radio-frequency (RF) communication may have a different meaning than for air-shower pulses.
This is because the typical duration of the signals is vastly different, being of the order of one to a few oscillation periods for the radio pulses of particle cascades and many oscillation periods in RF transmission or communication.
Despite these several advantages of power-like SNR definitions, it is not obvious whether this is sufficient to justify the use of power-like SNR units even in cases when the actual process or analysis is based on amplitudes (e.g., a voltage level used as trigger threshold). 
Therefore, the discussion regarding this point has not come to a final conclusion, but it seems recommendable to use power-like units whenever easily possible.

\subsection{Subtraction of Noise from the Signal}
When determining the signal-to-noise ratio, should the 'signal' $S$ be the true signal only, or should it include whatever quantity is obtained in the signal measurement, which usually contains the true signal and the noise? 
While it is often straightforward to subtract noise from the measured signal in many areas of physics, it is usually not the case for radio measurements. 
Even when using power-like quantities for signal and noise, the constructive/destructive interference in a specific event can be different from the average influence of noise (on average the power of signal and noise adds up).
Moreover, for low signals and when noise and signal are determined in different time intervals, the noise measured in a noise interval can be higher by chance than the signal measured in the signal interval, which would results in negative SNR values if noise is subtracted from the signal.

For these reasons, it is easier for experiments to use the measured signal including noise when determining the SNR. 
In the same way, it is easier in simulation studies to use the true simulated signal before adding noise when calculating the SNR. 
This situation results in differently defined SNRs for experimental measurements and simulation studies, which may be tolerable unless both are directly compared. 

While a common standard about whether or not noise should be subtracted from measurements or added to simulations will facilitate the understanding of SNR values, it would also have disadvantages. 
First, it requires processing measured signals before calculating the SNR, which is impossible in certain situations, e.g., when using the SNR for a threshold cut and only events above the cut are recorded or processed. 
Second, even in analysis where data are processed, it is not trivial to subtract the noise as the effect of the interference of noise and signal can depend on many factors, and possibly changes over time due to variations in the type and level of noise in many experiments. 
Therefore, also regarding this aspect, no final conclusion was achieved. 
For the time being, it is strongly encouraged for everyone to state clearly whether the 'signal' in any SNR definition includes noise or not.

\subsection{SNR of Background}
The naive expectation that the SNR of pure background (e.g., thermal or Galactic noise) is on average one, or zero if the noise is subtracted, is unfortunately incorrect. 
When determining the SNR of radio waveforms of pure background, without any signal in it, the SNR value varies due to fluctuations of noise.
Furthermore, the mean SNR value depends not only on the formula used for the SNR, but also on the methods used to determine noise and signal as well as on the type of background. 
Unless special care is taken, the mean SNR of noise will be very different from 1.
This often is a surprise for people new to the field, although it has been noted several times already: in LOPES \cite{Apel:2021oco}, it was a significant effort to change the determination of the noise level $N$ such that it is consistent with the way the signal $S$ is determined and that the SNR of pure noise is on average one \cite{LOPES:2010xob}. 
However, the approach chosen there requires sufficiently long traces for each event, and cannot easily be transferred to all experiments. 

Therefore, many experiments have to live with the situation that the SNR of noise is much larger than 1, hampering an intuitive interpretation. In an earlier version of the AERA \cite{vandenBerg:2013nza} analysis, the mean SNR was $5.0$ \cite{Glaser:2017ctn}, using the same band of $30-80\,$MHz and the same method to calculate SNR in Tunka-Rex \cite{Bezyazeekov:2015rpa}, the mean was $5.8$ in Tunka-Rex \cite{Hiller:2016wxe}: $SNR = S^2/N^2$ with the signal $S$ being the peak instantaneous amplitude in a signal search window of the electric-field trace and the noise $N$ being the RMS in a noise window of the same trace.

\begin{figure}[t]
\centering
\includegraphics[width=0.53\linewidth]{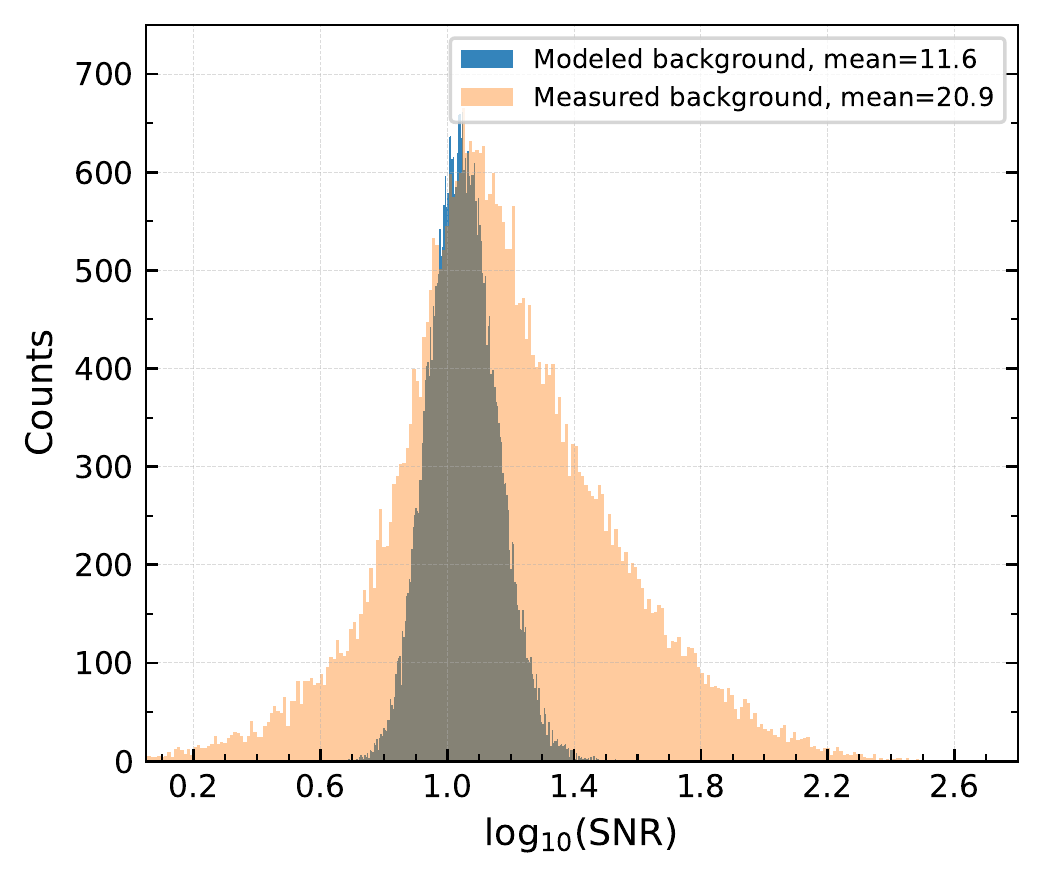}
\caption{SNR of background measured by SKALA v2 antennas at the South Pole in orange and background modeled by Cane Galactic + thermal noise in blue (appears gray due to the overlay). Both the width and mean of the SNR distribution are significantly higher for the measured background although the same definition and method are used to determine the SNR.}
\label{fig:differentBackgrounds}
\end{figure}

\begin{figure}[t]
\centering
\includegraphics[width=0.47\linewidth]{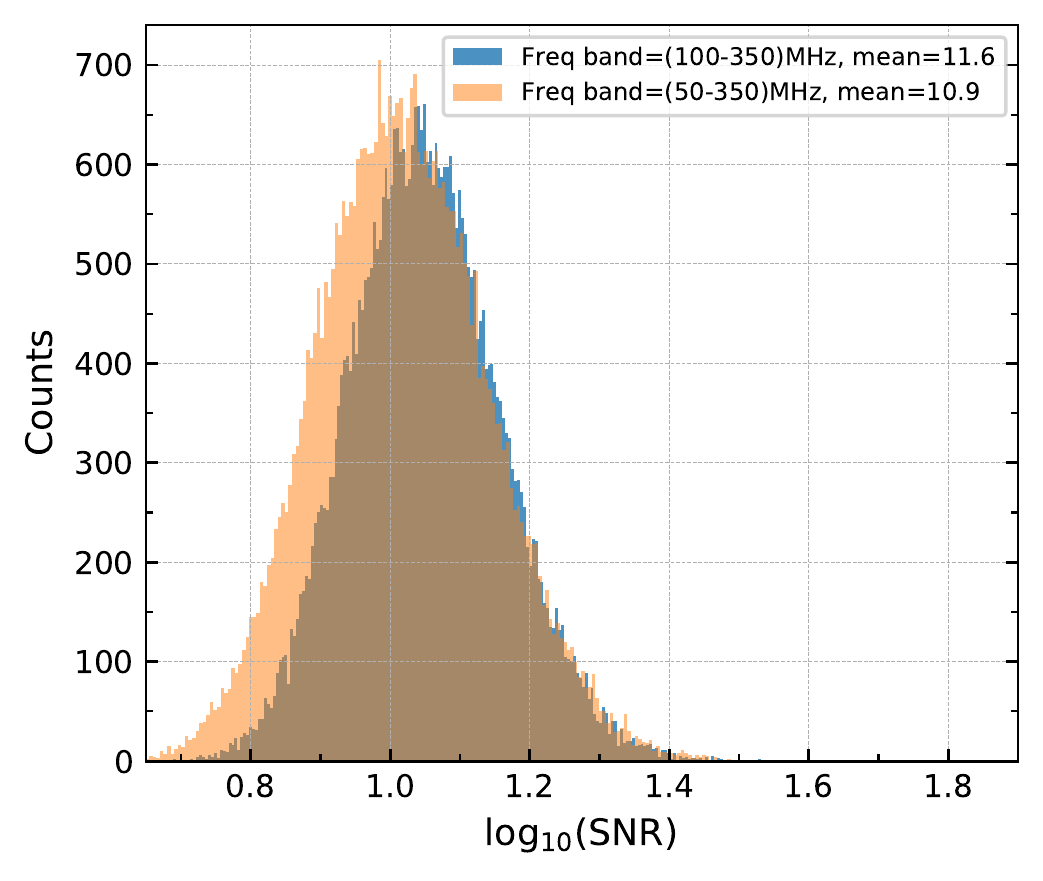}
\hfill
\includegraphics[width=0.47\linewidth]{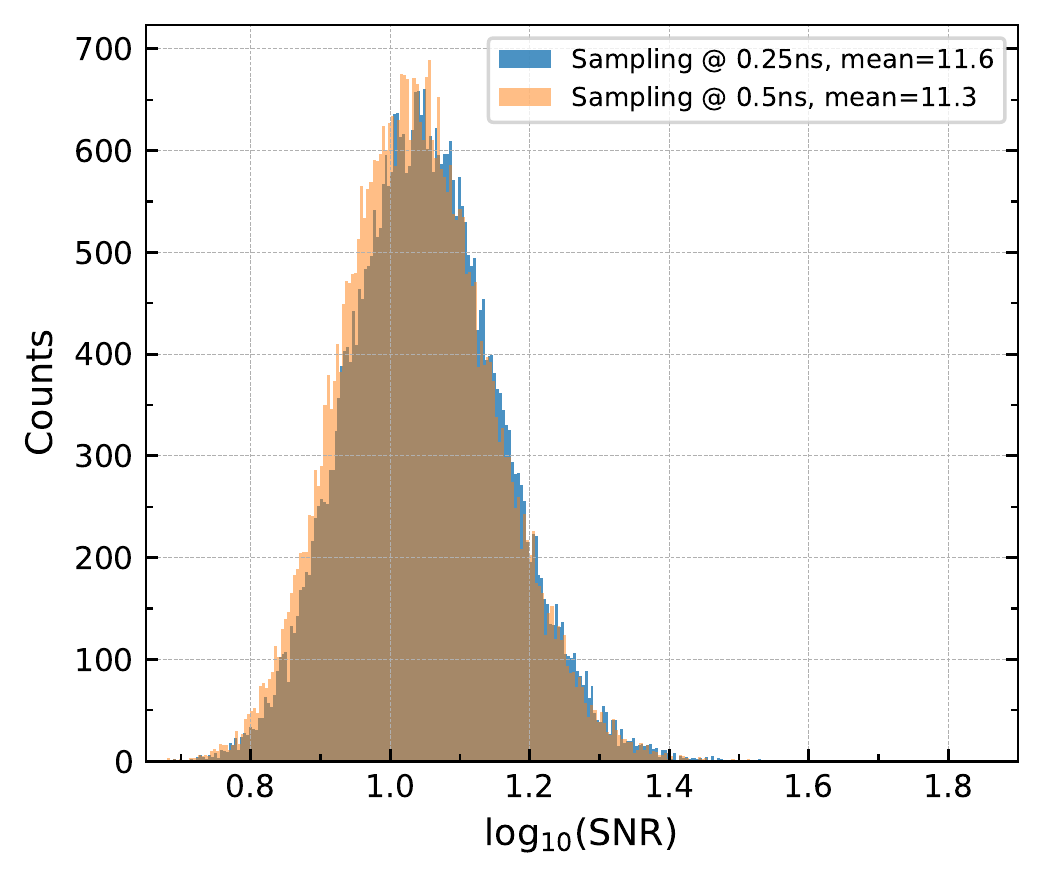}
\includegraphics[width=0.53\linewidth]{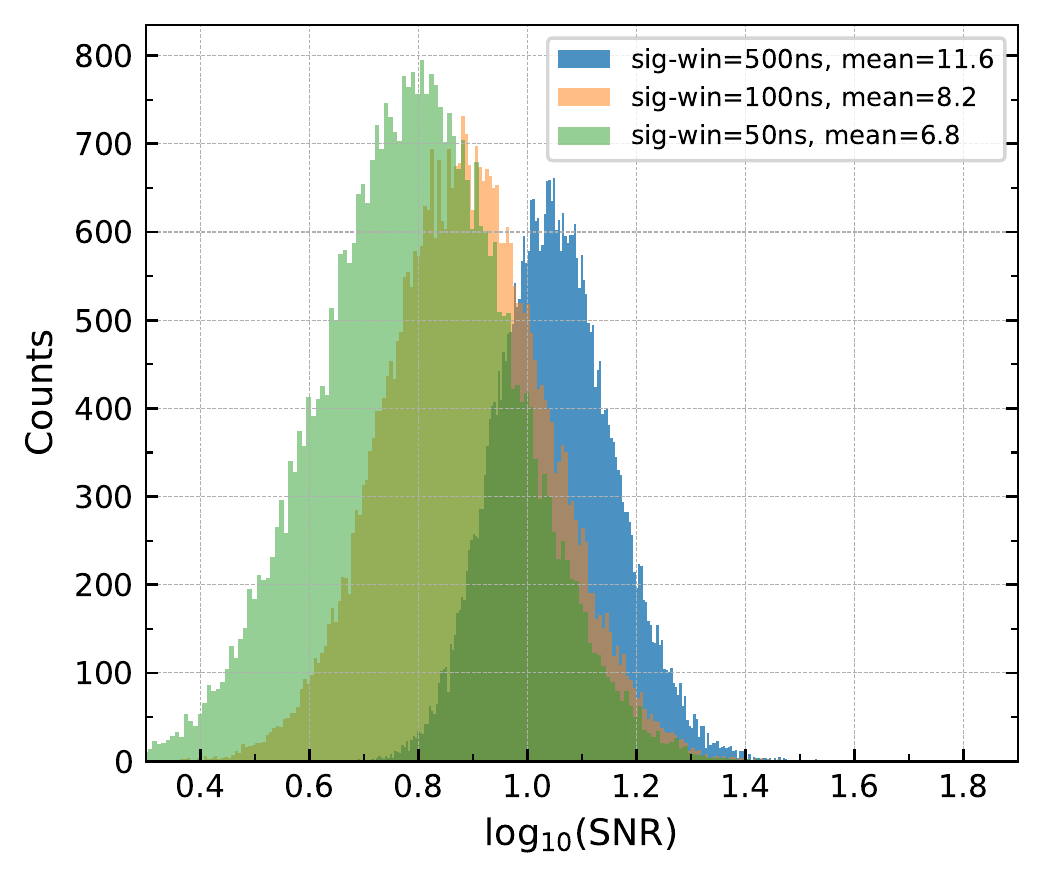}
\caption{Tested dependencies of the SNR distribution of modeled background. Top: there is a weak dependence on the frequency band (left), and on the upsampling frequency of the trace (right). Bottom: the mean SNR of pure background strongly depends on the size of the signal window ($50\,$ns vs.~$100\,$ns vs.~$500\,$ns), i.e., the time interval used to search for the maximum amplitude. The size of the noise window is $450\,$ns in all cases.}
\label{fig:differentBandsAndSampling}
\end{figure}

We have thus made a few exemplary investigations regarding the SNR of pure background.
First, we studied whether the assumptions hold that the SNR of pure background, if not one, at least is a constant depending only on the definition of the SNR and the method used to determine the SNR.
In one case, we used generated background with a model containing Cane Galactic noise and thermal noise, and in the other case background measured by a SKALA v2 antenna \cite{SKALAv2} of a prototype station at the South Pole \cite{IceCube:2021epf}.
In both cases, the traces were filtered to $100-350\,$MHz, and the SNR was determined as a power-like quantity by squaring amplitude-like quantities: the maximum amplitude in the time series of one antenna channel as signal and the RMS of the same time series as noise, where different time intervals were used for the signal and noise measurements \cite{AbdulARENA2022}.
The SNR definition using the squared peak amplitude as signal and the squared RMS of the trace as noise measure is very similar to the one which yielded mean SNR values of pure noise of $5.0$ and $5.8$ in the AERA and Tunka-Rex analyses cited above.
Nonetheless, the mean SNR value of pure noise is two to three times larger in this example, and also depends on whether the modelled or measured background sample is used (Fig.~\ref{fig:differentBackgrounds}).
This example shows that even for the same definition of SNR and when using the same methods to calculate it, the mean SNR of pure noise and, thus, the interpretation of SNR values can differ.

For the same definition of SNR and the same data set, the SNR value also depends on some aspects of the method on how the SNR is calculated. 
We have exemplarily checked a few possible dependencies for this particular definition of the SNR (Fig.~\ref{fig:differentBandsAndSampling}); more detailed investigations will be needed to obtain a more complete picture. 
In particular, we have found that the mean SNR of background varies slightly when changing the frequency band or when changing the sampling frequency. 
However, there is a strong dependence of the mean SNR of background on the size of the signal window, which is plausible since the chance of upward fluctuations of noise increases with the size of the time interval used to search for the maximum. 
Although this study is far from being exhaustive about the possible dependencies, it illustrates that the SNR has a flexible scale depending on the situation of the experiment and the way that that data are processed and the SNR determined.

Consequently, it is useful to state the SNR of pure background as a reference. 
For instance, an SNR value of 5 at LOPES corresponds to a large signal that stands out clearly from the noise (as the SNR of noise at LOPES was 1), however, in certain AERA or Tunka-Rex analyses, it would be considered a signal that is hardly distinguishable from the noise. 
Therefore, one suggestion that was discussed is to use $1/<SNR(\rm{background})>$ as a normalization constant, which would make the SNR values of different experiments and analyses more comparable. 
However, as no agreement has yet been reached on any standardization, we instead recommend to at least state the mean SNR value of the background as a reference, in order to facilitate the interpretation of the SNR scale.

\section{Conclusion}
The signal-to-noise ratio (SNR) is highly dependent on various aspects, and the scale can easily vary by a factor of 10 or more depending on the formula used to calculate the SNR, depending on some of the details of the data processing, such as the size of the signal time window, and depending on the actual background.

Therefore, the definition of SNR used as well as the method of calculation should be clearly stated in any publication. 
This includes in particular: \\
\begin{itemize}
    \item the formula used to calculate the SNR,
    \item whether amplitude- or power-like quantities are used for signal and noise,
    \item a clarification whether the 'signal' includes noise or whether the noise is subtracted,
    \item details on the method of calculation, such as the lengths of signal and noise windows,
    \item and the mean SNR value obtained when determining the SNR of pure background.
\end{itemize}

Of course, these clarifications can be done by citing an appropriate reference with detailed explanations, and there is no need to blow up the size of radio-related publications. 
Even though we have not yet agreed on a common standard, placing a remark in each publication on how the SNR is calculated will hopefully reduce misunderstandings, help newcomers in our community, and serve the important scientific standard of reproducibility.

\bibliographystyle{JHEP}
\bibliography{references}

\providecommand{\href}[2]{#2}\begingroup\raggedright\begin{thebibliography}{10}

\bibitem{Huege:2016veh}
T.~Huege, \emph{{Radio detection of cosmic ray air showers in the digital
  era}}, \href{https://doi.org/10.1016/j.physrep.2016.02.001}{\emph{Phys.
  Rept.} {\bfseries 620} (2016) 1}
  [\href{https://arxiv.org/abs/1601.07426}{{\ttfamily 1601.07426}}].

\bibitem{Schroder:2016hrv}
F.G.~Schr\"oder, \emph{{Radio detection of Cosmic-Ray Air Showers and
  High-Energy Neutrinos}},
  \href{https://doi.org/10.1016/j.ppnp.2016.12.002}{\emph{Prog. Part. Nucl.
  Phys.} {\bfseries 93} (2017) 1}
  [\href{https://arxiv.org/abs/1607.08781}{{\ttfamily 1607.08781}}].

\bibitem{Connolly:2016pqr}
A.L.~Connolly and A.G.~Vieregg, \emph{{Radio Detection of High Energy
  Neutrinos}},  in \emph{{Neutrino Astronomy: Current Status, Future
  Prospects}}, {World Scientific} (2017),
  \href{https://doi.org/{10.1142/9789814759410\_0015}}{DOI}
  [\href{https://arxiv.org/abs/1607.08232}{{\ttfamily 1607.08232}}].

\bibitem{BrayReview2016}
{J.D.~Bray}, \emph{{The sensitivity of past and near-future lunar radio
  experiments to ultra-high-energy cosmic rays and neutrinos}},
  {\emph{Astropart. Phys.} {\bfseries 77} (2016) 1}.

\bibitem{Schroder:2010ffv}
F.G.~Schr\"oder, \emph{{Instruments and Methods for the Radio Detection of High
  Energy Cosmic Rays}}, Ph.D. thesis, Karlsruhe Institute of Technology (KIT),
  2012.
\newblock 10.5445/IR/1000022313 and 10.1007/978-3-642-33660-7.

\bibitem{Glaser:2017ctn}
J.C.~Glaser, \emph{{Absolute energy calibration of the Pierre Auger observatory
  using radio emission of extensive air showers}}, Ph.D. thesis, RWTH Aachen,
  2017.
\newblock 10.18154/RWTH-2017-02960.

\bibitem{PierreAuger:2016vya}
{\scshape Pierre Auger} collaboration, \emph{{Measurement of the Radiation
  Energy in the Radio Signal of Extensive Air Showers as a Universal Estimator
  of Cosmic-Ray Energy}},
  \href{https://doi.org/10.1103/PhysRevLett.116.241101}{\emph{Phys. Rev. Lett.}
  {\bfseries 116} (2016) 241101}
  [\href{https://arxiv.org/abs/1605.02564}{{\ttfamily 1605.02564}}].

\bibitem{Apel:2021oco}
{\scshape LOPES} collaboration, \emph{{Final results of the LOPES radio
  interferometer for cosmic-ray air showers}},
  \href{https://doi.org/10.1140/epjc/s10052-021-08912-4}{\emph{Eur. Phys. J. C}
  {\bfseries 81} (2021) 176}
  [\href{https://arxiv.org/abs/2102.03928}{{\ttfamily 2102.03928}}].

\bibitem{LOPES:2010xob}
{\scshape LOPES} collaboration, \emph{{On noise treatment in radio measurements
  of cosmic ray air showers}},
  \href{https://doi.org/10.1016/j.nima.2010.11.009}{\emph{Nucl. Instrum. Meth.
  A} {\bfseries 662} (2012) S238}
  [\href{https://arxiv.org/abs/1009.3444}{{\ttfamily 1009.3444}}].

\bibitem{vandenBerg:2013nza}
{\scshape Pierre Auger} collaboration, \emph{{Results from and prospects for
  the Auger Engineering Radio Array}},
  \href{https://doi.org/10.1051/epjconf/20135308006}{\emph{EPJ Web Conf.}
  {\bfseries 53} (2013) 08006}.

\bibitem{Bezyazeekov:2015rpa}
{P. A. Bezyazeekov et al. - Tunka-Rex Collaboration}, \emph{{Measurement of
  cosmic-ray air showers with the Tunka Radio Extension (Tunka-Rex)}},
  \href{https://doi.org/10.1016/j.nima.2015.08.061}{\emph{Nucl. Instrum. Meth.
  A} {\bfseries 802} (2015) 89}
  [\href{https://arxiv.org/abs/1509.08624}{{\ttfamily 1509.08624}}].

\bibitem{Hiller:2016wxe}
R.~Hiller, \emph{{Radio measurements for determining the energy scale of cosmic
  rays}}, Ph.D. thesis, Karlsruhe Institute of Technology (KIT), 2016.
\newblock 10.5445/IR/1000053456.

\bibitem{SKALAv2}
E.~de~Lera~Acedo et~al., \emph{Evolution of skala (skala-2), the log-periodic
  array antenna for the ska-low instrument},
  \href{https://doi.org/10.1109/ICEAA.2015.7297231}{\emph{2015 International
  Conference on Electromagnetics in Advanced Applications (ICEAA)} (2015) 839}.

\bibitem{IceCube:2021epf}
{\scshape IceCube} collaboration, \emph{{First air-shower measurements with the
  prototype station of the IceCube surface enhancement}},
  \href{https://doi.org/10.22323/1.395.0314}{\emph{PoS} {\bfseries ICRC2021}
  (2021) 314} [\href{https://arxiv.org/abs/2107.08750}{{\ttfamily
  2107.08750}}].

\bibitem{AbdulARENA2022}
{\scshape IceCube} collaboration, \emph{{Deep Learning for the Classification
  and Recovery of Cosmic-Ray Radio Signals against Background Measured at the
  South Pole}}, \href{https://doi.org/10.22323/1.424.0012}{\emph{PoS}
  {\bfseries ARENA2022} (2023) 012}.

\end{thebibliography}\endgroup
\vspace{5mm}
\noindent
\footnotesize{This project has received funding from the European Research Council (ERC) under the European Union's Horizon 2020 research and innovation programme (grant agreement No 802729).}

\end{document}